\documentclass[%
aip, 
apl,
preprint,
superscriptaddress,
nofootinbib,
 amsmath,amssymb,
floatfix,
]{revtex4-1}

\usepackage{graphicx}
\usepackage{dcolumn}
\usepackage{bm}
\usepackage{color}
\usepackage{gensymb}
\usepackage[dvipsnames]{xcolor}

\begin{document}

\title{The integration of photonic crystal waveguides with atom arrays in optical tweezers}

\author{Xingsheng Luan}
\author{Jean-Baptiste B\'eguin}
\affiliation{Norman Bridge Laboratory of Physics, California Institute of Technology, Pasadena, California 91125, USA}
\author{Alex P. Burgers}
\affiliation{Norman Bridge Laboratory of Physics, California Institute of Technology, Pasadena, California 91125, USA}
\address{Present Address: Department of Electrical Engineering, Princeton University, Princeton, New Jersey 08540, USA}
\author{Zhongzhong Qin}
\affiliation{Norman Bridge Laboratory of Physics, California Institute of Technology, Pasadena, California 91125, USA}
\affiliation{State Key Laboratory of Quantum Optics and Quantum Optics Devices, Institute of Opto-Electronics, Shanxi University, 030006 Taiyuan, China}
\author{Su-Peng Yu}
\affiliation{Norman Bridge Laboratory of Physics, California Institute of Technology, Pasadena, California 91125, USA}
\affiliation{Present Address: Time and Frequency Division, NIST, 385 Broadway, Boulder, Colorado 80305, USA}
\author{H. J. Kimble}
\affiliation{Norman Bridge Laboratory of Physics, California Institute of Technology, Pasadena, California 91125, USA}
\affiliation{Corresponding author: hjkimble@caltech.edu}

\date{\today}

\begin{abstract}
Integrating nanophotonics and cold atoms has drawn increasing interest in recent years due to diverse applications in quantum information science and the exploration of quantum many-body physics. For example,
dispersion-engineered photonic crystal waveguides (PCWs) permit not only stable trapping and probing of ultracold neutral atoms via interactions with guided-mode light, but also the possibility to explore the physics of
strong, photon-mediated interactions between atoms, as well as atom-mediated interactions between photons. While diverse theoretical opportunities involving atoms and photons in 1-D and 2-D nanophotonic lattices have 
been analyzed, a grand challenge remains the experimental integration of PCWs with ultracold atoms. Here we describe an advanced apparatus that overcomes several significant barriers to current experimental progress with the goal of achieving strong quantum interactions of light and matter by way of single-atom tweezer arrays strongly coupled to photons in 1-D and 2-D PCWs. Principal technical advances relate to efficient free-space coupling of light to and from guided modes of PCWs, silicate bonding of silicon chips within small glass vacuum cells, and deterministic, mechanical delivery of single-atom tweezer arrays to the near fields of photonic crystal waveguides.
\end{abstract}

\maketitle

\section{Introduction \label{sec:intro}}

The research described in this manuscript attempts to create novel paradigms for strong quantum interactions of light and matter by way of single atoms and photons in nanoscopic dielectric lattices. Nanophotonic structures offer the intriguing possibility to control  interactions between atoms and photons by engineering the medium properties through which they interact\cite{Kimble:18,Lodahl:17}. Opportunities beyond conventional Quantum Optics thereby emerge for unconventional quantum phases and phenomena for atoms and photons in one and two-dimensional nanophotonic lattices\cite{Gonzalez-Tudela:15,Douglas:15,Yu:19}. The research is inherently multidisciplinary, spanning across nanophotonics, atomic physics, quantum optics, and condensed matter physics.

Beyond the advances reported here, this general area has diverse implementations for Quantum Information Science, including the realization of complex quantum networks\cite{Kimble:08} and the exploration of quantum many-body physics with atoms and photons\cite{Hung:16}. Further avenues of interest are the investigation of quantum metrology and long distance quantum communication\cite{Covey:19} combined with the integrated functionality of nanophotonics and atoms. As a comparison, solid-state emitters coupled to nanophotonic structures~\cite{Sipahigil:16,Dibos:18,Zhong:18} provide a complementary route to some of the physics described here. However, these systems exhibit inhomogeneous broadening which can make the coupling of even two such emitters in a single nanophotonic structure a challenging experimental task~\cite{Evans:20} and they are not, in their current form, designed to generate controllable interactions across a large system of emitters as has been demonstrated with numerous atomic physics platforms.

While exciting theoretical opportunities of atoms coupled to nanophotonics have emerged, this research only moves forward in the laboratory by advancing nanophotonic device fabrication and by integrating these novel devices into the realm of ultracold atoms. Important experimental milestones have been reached~\cite{Thompson:13,Tiecke:14,Goban:14,Goban:15,Hood:16,Samutpraphoot:20}, but generally laboratory progress has lagged theory in combining ultracold atoms and novel nanophotonic devices. As illustrated in Figure \ref{fig:1D2D}, a grand challenge for experiments in this new field is the realization of atomic arrays with high fractional filling of single atoms into unit cells of 1D and 2D lattices\cite{Kimble:18,Lodahl:17}. In this manuscript we describe an apparatus that provides several significant advances relative to prior technical capabilities, that are summarized as follows: 

\begin{figure}[htbp]
\centering
\vspace{-5mm}
\includegraphics[width=0.7\linewidth]{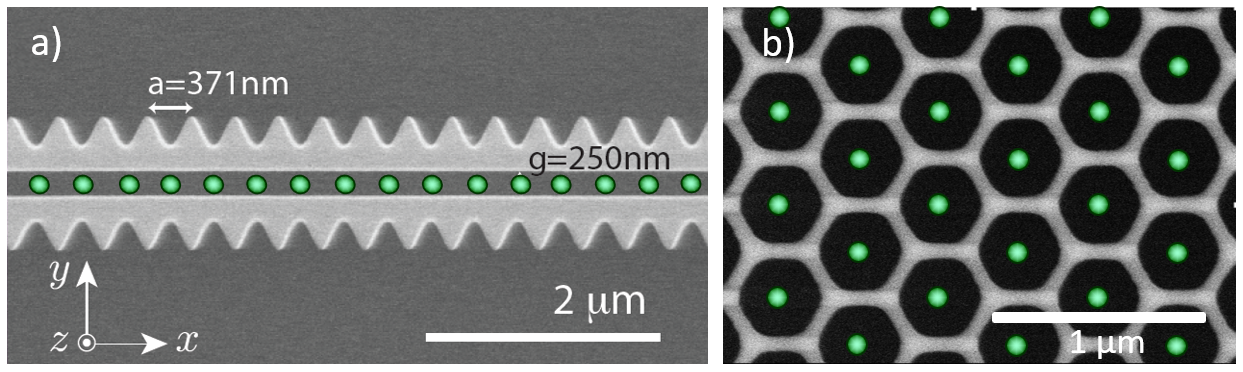}
\vspace{-5mm}
\caption{Schematic for building nanoscopic atomic arrays with one atom per unit cell in (a) 1-D alligator photonic crystal waveguide (APCW) \cite{Goban:14,Yu:14} and (b) 2-D honeycomb lattice PCW \cite{Yu:19}, respectively. In both cases the Silicon Nitride structures (gray) are suspended above an underlying Silicon substrate (dark background). Green spheres represent the yet to be achieved single atoms trapped by optical forces within unit cells of the photonic crystal structures\cite{Hung:13,Burgers:19}}
\label{fig:1D2D}
\end{figure}

\begin{figure}[t]
\centering
\includegraphics[width=\linewidth]{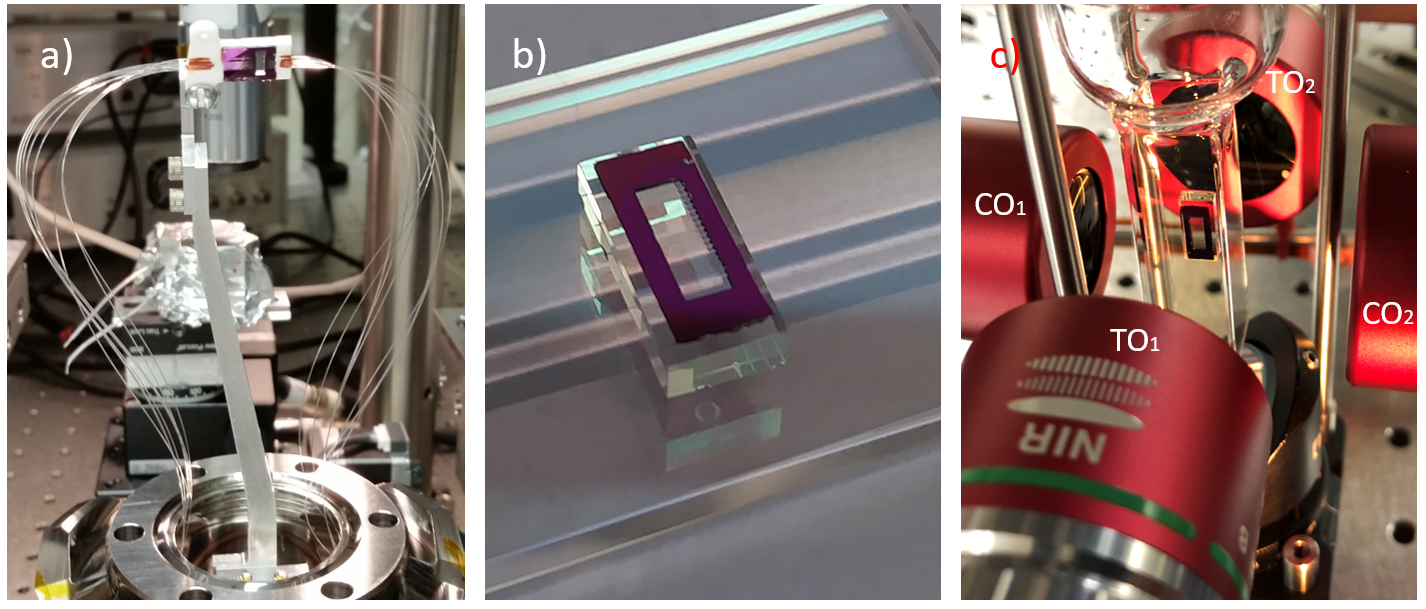}
\caption{The old (a) and new (b,c) ways of integrating nanophotonic chips with cold-atom vacuum systems; (a) scale for old system is set by 2.75'' flange to connect to conventional 6'' diameter vacuum chamber with chip (top of figure) then centered in chamber. For more details, see Chapter 6 and Figure 6.4 in Ref. ~\onlinecite{Munizthesis:17}. (b) New system with Silicon chip of 4 mm width mounted on a SiO$_2$ `optical table' of dimensions 5$\times$11$\times$2 mm. For more details, refer to Section \ref{sec:sb}. (c) The assembled system in a SiO$_2$ glass cell of internal dimension 1$\times$1$\times$4.5 cm (rectangular part), surrounded by two coupling objectives (CO$_1$ and CO$_2$ with $N.A.=0.4$) for free-space optical coupling to photonic crystal waveguides and two tweezer objectives (TO$_1$ and TO$_2$ with $N.A.=0.4$) for generating optical tweezer traps and imaging\cite{Beguin:20}.}
\label{fig:old_new}
\end{figure}

1. Silicate bonding -- We have previously used large manipulators inside conventional stainless steel chambers for mounting our Silicon chips illustrated in Figure \ref{fig:old_new}a. This method posed several limitations in our previous experiments\cite{Goban:14,Goban:15,Burgers:19}, including mechanical and thermal instability, complexity in fiber alignment and assembly, and limitations on baking temperature and out-gasing due to various epoxy resins and bonding agents used for ``gluing" optical fibers to Silicon chips and chips to vacuum mounting hardware. Following discussions with Jun Ye and John Hall at JILA, we have developed a new platform to mount our chips in vacuum. As shown in Figure \ref{fig:old_new}b and also described in details in Section \ref{sec:sb}, we now bond a Si chip to an SiO$_2$ substrate by way of silicate bonding that we have developed in our group (with significant input from a LIGO research team at Caltech).  This bonding technique has extremely low out-gassing properties compared to our previous chip mounting configurations and enables high temperature baking of our vacuum cell for ultra-high vacuum (UHV) operation.

2. A new generation of PCWs -- We have developed optical chips that eliminate fiber optics within the vacuum chamber, achieve more efficient coupling of light into and out of our PCWs, increase power handling capabilities by twenty fold to facilitate long-lived guided-mode optical traps\cite{Hung:13}, and enable high-temperature baking for improved atom trapping times\cite{Yuthesis:17}. All of these goals have been met by way of the design and fabrication of devices that utilize free-space optical coupling whereby input laser light is coupled from outside the vacuum chamber directly into individual PCWs. As described in Section \ref{sec:ycoupler}, we have removed the need for in-vacuum fibers by designing and fabricating a new Y-coupler at the terminating ends. The Silicon chip also has a significantly reduced footprint necessitated by having the terminating Y-couplers much closer to the edges of the Silicon chip to allow free-space coupling for numerical apertures (N.A.) $\sim 0.1-0.2$ of the new Y-couplers\cite{Yuthesis:17}. 

3. A new laboratory -- Free-space coupling and silicate bounding have enabled the construction of a new laboratory in the Quantum Optics Group at Caltech, which is built around a vacuum envelope reduced in size by approximately a factor of $10^2$ to reach a volume $\sim 10$ cm$^3$ with unprecedented optical access relative to our prior chambers, as shown in Figure \ref{fig:old_new}c. We aim to achieve nanoscopic lattices of atoms that are assembled deterministically with arrays of single-atom tweezers and that are coupled to guided modes (GMs) of PCWs for efficient atom-photon coupling along the PCW and of GM photons to and from free-space. As described in Section \ref{sec:atom}, our experiment is in the spirit of recent worldwide advances with free-space tweezer arrays\cite{Kaufman:12,ThompsonPRL:13,Endres:16,Barredo:16} but adds the significant complexity of assembling such atomic arrays near the surfaces of 1D and 2D PCWs.

A general summary of our advances is provided by Figure \ref{fig:old_new}, which shows our `old'\cite{Yuthesis:17,Munizthesis:17,McClung:17} and `new'\cite{Beguin:20} systems side by side. Of course, small glass cells with volume $\sim 1$ cm$^3$ for various optical trapping schemes have been employed by various groups for many years\cite{Anderson95}. But to our knowledge, no group has succeeded to implement a compact setup as in Figure \ref{fig:old_new}(b, c) when the difficult constraints of localization of atoms near PCWs have been part of the setup\cite{Beguin:20}.

\section{Silicate bonding \label{sec:sb}}

\begin{figure}[htbp]
\centering
\includegraphics[width=\linewidth]{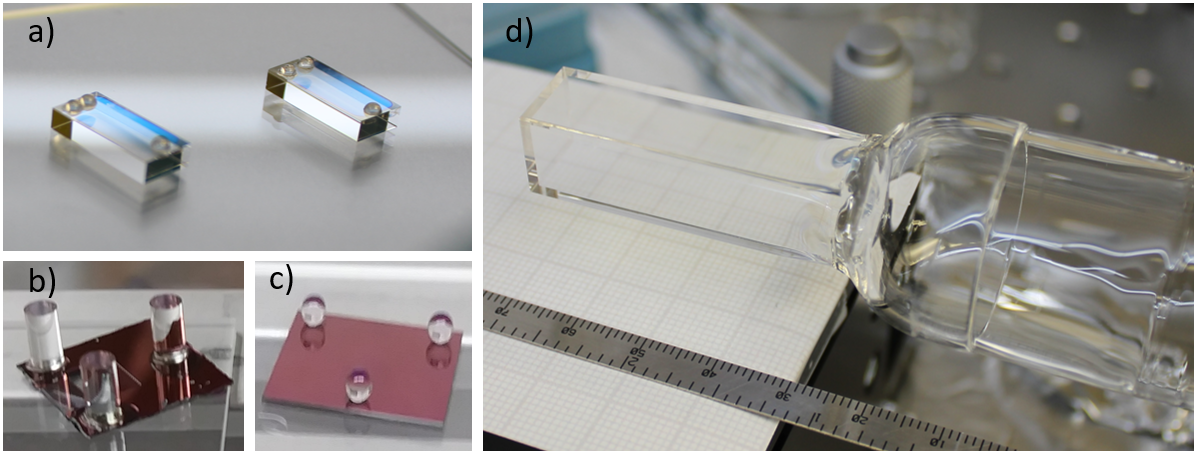}
\caption{Photographs of (a) SiO$_2$ polished rectangular tables (5$\times$11$\times$2 mm) with three SiO$_2$ hemispheres silicate-bonded onto the AR-coated table surface, (b) Three parallel SiO$_2$ cylindrical rods with diameter $1$ mm and length $2$ mm individually bonded on flat cross-section to an APCW Si chip, (c) Three SiO$_2$ spheres (diameter $1.5$ mm) bonded to a blank 1$\times$1 cm square Silicon chip, (d)} SiO$_2$ glass chamber with external AR-coating on all 5 surfaces of the rectangular glass cell with 1.25 mm wall thickness, internal square cross-section 1$\times$1 cm and length 4.5 cm; a Quartz to Pyrex graded seal (internal diameter $2.54$ cm) is fused via a cup (for coating protection) to the rectangular cell, for chamber bake-out temperatures up to 400 K.
\label{fig:1}
\end{figure}

In previous atom-nanophotonic experiments\cite{Goban:14,Goban:15,Hood:16,Burgers:19} in the Quantum Optics Group at Caltech, the nanophotonic chips were held inside a stainless vacuum chamber by a long mechanic manipulator ($\sim$ 10 cm) with input-output fibers pre-aligned and glued in the V-groove on chip\cite{Yu:14}, as shown in Figure \ref{fig:old_new}a. This posed three major limitations in our previous experiments: 1) The long mechanical arm did not have sufficient mechanical stability, thereby limiting the ability for precise positioning of atoms on nanophotonic structures and dissipation of guided mode heating power. 2) Fiber coupling of light into and out of the chip involved cumbersome fiber pre-alignment and gluing outside the vacuum chamber, and the number of devices (8 APCWs) that could be coupled was limited by failure probability and the number of vacuum feedthroughs available (8 input plus 8 output fibers). Furthermore, once the chip was inside the vacuum chamber, the coupling efficiency could not be further adjusted or optimized. 3) The usage of UHV compatible epoxy for gluing fibers and the chip prohibited the possibility of significantly baking the entire vacuum chamber. As a result, the typical lifetime for atoms trapped near ($\sim300$ $\mu$m) the chip was limited to be $\lesssim100$ ms\cite{Goban:15}.

Here, by adapting the silicate bonding method~\cite{Veggel:14} whose reliability was demonstrated in NASA and ESA astronomical satellite missions (e.g. Gravity Probe B and The LISA Pathfinder) and current LIGO instruments, we are now able to overcome these limitations by mounting the nanophotonic chip inside a glass cell with small footprint that is compatible with free-space coupling from microscope objectives outside the vacuum cell. This largely eliminates the relative motion between chip and vacuum chamber, and also the need of all optical fibers within the vacuum envelope.

The hydroxide catalysis bonding method of Ref.~\onlinecite{Gwo:01} involves strong chemical bonds between oxidizable materials such as SiO$_2$ and Silicon. Such chemical bonds can be formed at room-temperature. Optically, silicate bonding provides a transparent bond, the refractive index of which between two SiO$_2$ surfaces, converges to the index of SiO$_2$~\cite{Mangano:17} thereby minimizing Fresnel reflections from the bonded surfaces and allowing low-loss optical transmission through the bond. AR-coated glass surfaces can also be silicate bonded if terminal layers of SiO$_2$ are deposited on the surfaces to be bonded. Silicate bonding allows UHV operation, which is an important requirement for research involving trapped cold atoms near surfaces inside a vacuum chamber. The operating temperature for components secured by silicate bonding ranges from cryogenic to beyond typical bake-out temperatures for UHV chambers (i.e., $300-400$ K). Because of the UHV compatibility of silicate bonding and the small footprint of the glass cell, a vacuum pressure of $\sim10^{-11}$ Torr is achieved after the first baking of the entire vacuum setup, as compared to $\sim10^{-9}$ Torr in previous work.

\begin{figure}[htbp]
\centering
\includegraphics[width=\linewidth]{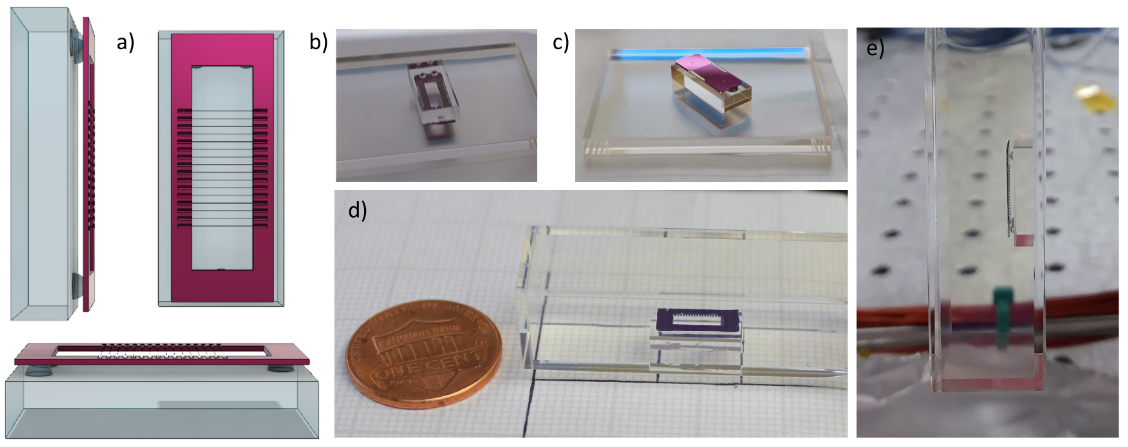}
\caption{(a) 3D model drawings of Fused Silica glass table assembly with Silicon chip with three different perspective views. (b,c,d,e) Photographs of table assembly during construction and inside coated vacuum glass cell. (b) The weight of the glass table with hemispheres is used during bonding to the Si-chip frame. (c) Example photograph of a window-less Si-chip bonded to the glass table. (d,e) The small chip-table assembly is finally bonded onto one inner wall of the rectangular glass cell.}
\label{fig:1b}
\end{figure}

Turning then to the steps for achieving a mounted Silicon chip by way of Silicate bonding, we show in Figures \ref{fig:1} and \ref{fig:1b} photographs of various stages of the sequence. Figure \ref{fig:1}a shows the glass tables upon which 200 $\mu$m thick (4$\times$9 mm) Silicon chips will be bonded via three effective contact surfaces as in (b-c) using rods and spheres as well as hemispheres in (a). Here we consider the case of hemispheres for which three SiO$_2$ hemispheres are first bonded onto a SiO$_2$ rectangular prism table. The curved caps of the hemispheres are then flat polished to better than $\lambda/10$ over a circular area of diameter $\sim 0.8$ mm, defining a precision plane (to within $100$ nm) for next bonding the Silicon chip to the flat tops of the polished hemispheres and hence to the optical table. The table-chip assembly is then itself bonded to the inner wall of a precision Fused Quartz glass cell fabricated by Starna\cite{Starna} shown in Figure \ref{fig:1b}(e). It should be noted that the rectangular prism table is AR-coated on its outer side (i.e., facing into the glass cell), while the glass cell is AR-coated on its outer surface but not inner.

We recall that the nanophotonic structures are e-beam written into a 200 nm sacrificial layer of Silicon Nitride deposited on the 200 $\mu$m thin Silicon substrate\cite{Yu:14}, so that considerable care is required to avoid damage to the surface containing the devices. Without the use of additional optical elements, the bare divergence angle of the light emerging from the nano-waveguides ($N.A. \sim 0.15$, as in Ref.~\onlinecite{Yuthesis:17}) requires elevating the chip base from any surface and to position it in relation to the glass cell geometry to avoid clipping loss.

\section{The Y-coupler technology\label{sec:ycoupler}}

In this section, we present a description of the new ``Y-coupler'' design which provides efficient free-space coupling, minimal light scattering and better mechanical stability. This design extends the maximum power by roughly 20x beyond the failure power for our previous fiber butt-coupled devices \cite{Yu:14} which was a major limitation in our previous atom-nanophotonic experiments\cite{Goban:14,Goban:15,Burgers:19} for achieving long-lived guided-mode atom trapping at magic wavelengths\cite{Ye:08,Hung:13} as previously demonstrated in the optical nanofiber system\cite{Vetsch:10,Goban:12}.

\begin{figure}[htbp]
\centering
\fbox{\includegraphics[width=\linewidth]{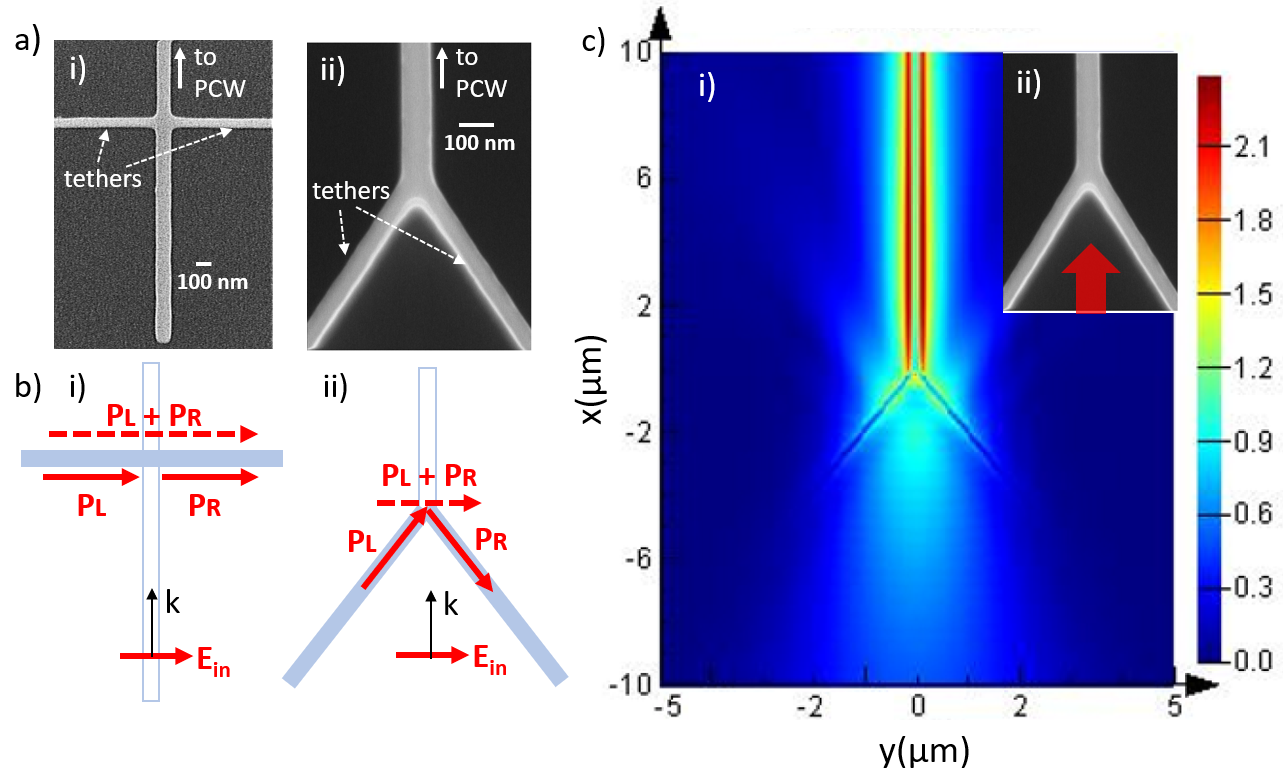}}
\caption{(a) SEM images of the coupler with conventional 90\degree tether termination (i) and the Y-coupler (ii)  (b) Vectorial illustration of the polarizabilities on tethers shows the working principle for the coupler with conventional tether termination (i) and the Y-coupler (ii). (c) FDTD simulation of wave propagation in Y coupler for an incident Gaussian beam propagating from the bottom to the top of the figure. Insert is a SEM image of device under simulation with the Gaussian mode input indicated as a red arrow.}
\label{fig:3}
\end{figure}

The chip design in this work is an adaptation of the system presented in Ref.~\onlinecite{Yu:14} to enable direct free-space coupling from an objective into the waveguides. The devices are fabricated from a 200 nm Silicon Nitride device layer, suspended from a 200 $\mu$m Silicon substrate. Precision grooves aligned to the waveguide device layer are etched into the substrate, to enable cleaving of the chip for clearance for free-space beam inputs and outputs. The absence of terminated optical fibers in the vicinity of the waveguide input coupler widens the design space available for coupler designs. Here, we present a Y-coupler design that simultaneously optimizes transmission, suppresses residual reflection, and provides mechanical stability\cite{Yuthesis:17}.\\ In order to mode-match the guided mode of a photonic waveguide to a Gaussian beam, we taper the waveguide width to $\simeq$ 130 nm, before terminating the waveguide and launching the mode. The terminated end of the suspended waveguide is affixed to the substrate through two $\leq$ 100 nm wide tethers. Conventionally, the tethers are simply arranged perpendicular to the waveguide \cite{Yu:14}, as shown in the SEM image in Figure~\ref{fig:3}a (i). Two issues arise from such tethering design. First, the tether pair is polarized by the guided mode light, creating scattering and back-reflection that are undesirable. Second, the tethers are perpendicular to the waveguide, therefore releasing the tensile stress on the waveguide, potentially making the tapered section of the waveguide mechanically pliable. At high optical power, the waveguide-tether junction is observed to produce fluctuating scattering intensity prior to mechanical failure, which we attribute to thermally-induced stress distributions causing physical movements of the junction. These pose a significant constraint for achieving guided mode traps at magic wavelengths for Cs with a typical power handling $\sim$ 10 mW, well beyond the $\sim$ 0.5 mW limit found for conventional couplers in Refs.~\onlinecite{Cohen:13,Yu:14}. 

To overcome this limitation, an unconventional Y-shaped termination of the free-standing end of the waveguide was designed in joint consideration of mechanical and optical properties, as shown in Figure \ref{fig:3}a (ii). The termination ends of the suspended waveguide need to be mechanically affixed to the substrate with tethers. By tilting the tethers away from the tapered waveguide, a controllable weak tensile stress can be maintained on the waveguide and tethers to make them mechanically robust. Optically, the tethers are tilted away from the electric field vector of the incoming mode, reducing the polarizability of each tether. This effect, in addition to the partial cancellation of the polarization vectors of the two tethers, reduce the total polarizability of the junction, therefore reducing the scattering loss due to the tethers, as shown in Figure \ref{fig:3}b (i) and (ii) with vectorial illustrations for conventional coupler with 90\degree~tethers and the Y-coupler, respectively. In practice, a tilt angle of 60\degree~from perpendicular was chosen from FDTD optimization\cite{Lumerical}. The simulated field pattern of a Gaussian beam incident on the junction from below is shown in Figure \ref{fig:3}c. Coupling efficiencies of a $1/e^2$ waist $w_0\sim2.5$ $\mu$m beam with different input polarizations to different couplers are calculated from FDTD simulations and are summarized in Table \ref{tab:coupler_fdtd}. It suggests that, the Y-coupler has a better performance than conventional couplers in terms of  Gaussian-beam-to-waveguide transmission (87\% vs. conventional 79\%) and reflection ($<0.1\%$ vs. conventional 2.7\%) for input polarization along y direction (TE), which is critical for quantum correlation measurements in nanophotonics\cite{Goban:14}. In experiment, we have measured a coupling efficiency up to $80\%$ for TE input using the Y-coupler design\cite{Beguin:20}. 

\begin{table}[htbp]
\centering
\caption{\bf FDTD simulated transmission and reflection efficiencies for different couplers}
\begin{tabular}{ccccc}
\hline
Type & TE Transmission &  TM Transmission & TE Reflection & TM Reflection \\
\hline
with $90\degree$  tethers & 79\% & 65\% & 2.7\% & 0.6\% \\
with Y-shape tethers & 87\% & 56\% & $<0.1\%$ & $<0.1\%$ \\
\hline
\end{tabular}
  \label{tab:coupler_fdtd}
\end{table}

\begin{figure}[htbp]
\includegraphics[width=\linewidth]{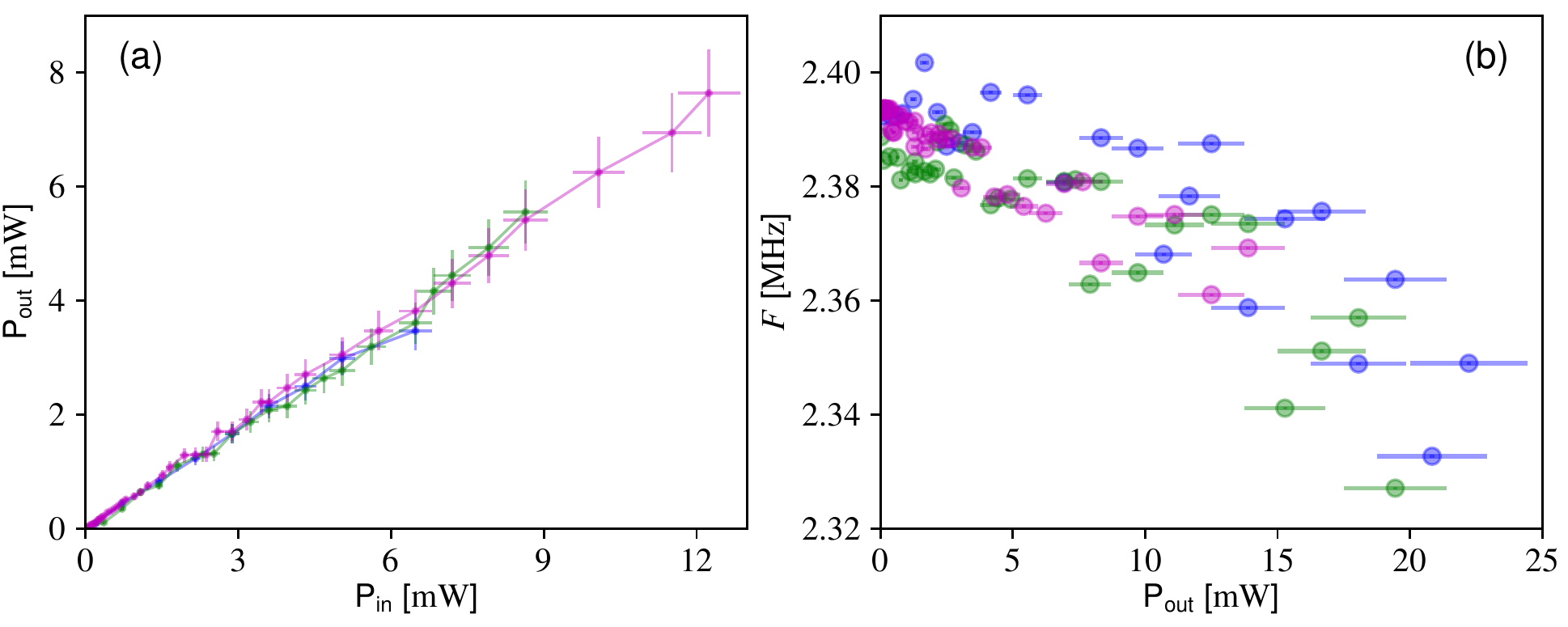}
\caption{(a) Measured output light power $P_{\text{out}}$ versus input light power $P_{\text{in}}$ for optical light field propagating in the TE mode with vacuum wavelength $935.7$ nm and chamber pressure of $\sim 3\times 10^{-10}$ Torr. (b) Fundamental mechanical frequency of the differential in-plane mechanical mode of APCW as a function of output light power. Different colors correspond to different nanophotonic devices on the same chip.}
\label{fig:3b}
\end{figure}

In Figure \ref{fig:3b}, we show measurement data asserting the power handling capability of the devices with the free-space coupling and Y-coupler design strategy. Figure~\ref{fig:3b}a shows the optical power transmitted by the device as a function of the input power for light propagating in the TE mode with a magic wavelength for Cs atoms at $935.7$ nm.  The new design allows a measured 20-fold increase (from $\sim 0.5$ mW to $\sim 10$ mW) in the maximum optical light power before breaking or irreversible damages as compared to our previous devices with the butt-coupler design\cite{Yu:14}. This should enable long-lived guided-mode atom traps by way of higher intensities required for larger atomic detunings, including for magic-wavelength traps \cite{Hung:13,Ye:08}.

While the detailed physics of device failure and plastic deformation is beyond the scope of this article, we show in Figure~\ref{fig:3b}b a measurement of the dependence of the mechanical frequency of the fundamental differential in-plane mode of motion of the APCW as a function of the output light power. Devices physically break at $P_{\text{out}} \sim 20$ mW.  At the very low powers, the quasi-linear decrease in frequency is compatible with a simple model of reversible thermal elongation of a highly stressed string. At the highest powers before failure, the frequency shift would amount to a relative physical elongation and equivalent strain of $\sim 0.04$, compatible with typical ratios of the yield strength of Silicon Nitride to its Young's modulus.

\section{Single atom trapping in a tweezer array near PCWs\label{sec:atom}}

\begin{figure}[htbp]
\centering
\fbox{\includegraphics[width=0.5\linewidth]{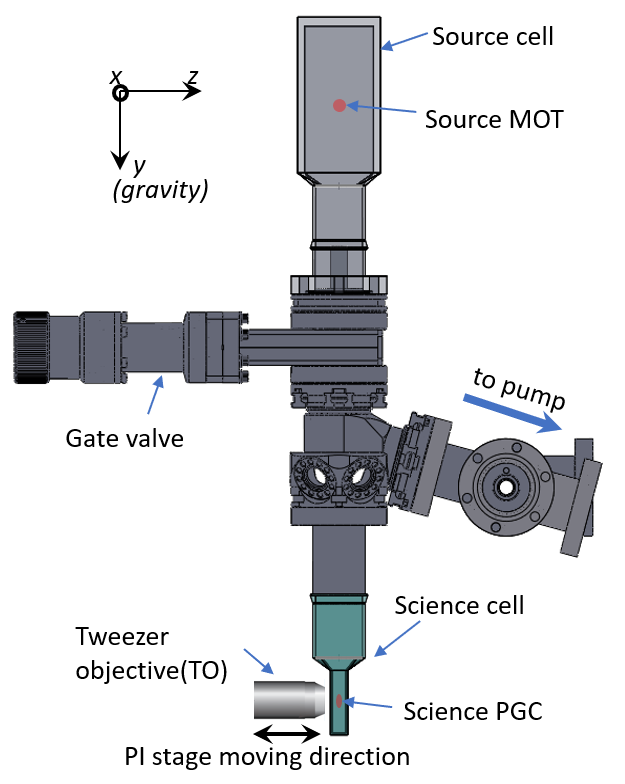}}
\caption{A simplified AutoCAD\textsuperscript{\textregistered} drawing of the experimental setup. The source MOT and science PGC cloud positions are also indicated with red circles. Falling atoms are delivered from source cell to science cell by the guiding of a blue-detuned donut-shaped beam (not shown).}
\label{fig:2_setup}
\end{figure}

One of the key challenges in atom-nanophotonic experiments is to achieve efficient loading of atoms into guided mode traps formed on the nanophotonic structures. Given the small guided mode trap volume and strong Casimir-Polder potential near the dielectric surfaces, it was shown in  Ref.~\onlinecite{Goban:14,Burgers:19} that direct loading of background Cs atoms into the APCW guided mode traps is very difficult. Trapping few atoms \textit{in a single trap} $\sim 130 $ nm above the APCW surface is demonstrated in Refs.~\onlinecite{Goban:15,Hood:16} by reflecting a waist $1/e^2$ $w_0\sim60$ $\mu$m dipole trap beam (the so-called side-illumination beam). However, the average trapped atom numbers $\bar{N}$ is restricted to $\sim 3$ atoms and the trap size along the APCW is $\sim10$ $\mu$m (along the x direction as in Figure \ref{fig:1D2D}a), corresponding to $\sim 27$ unit cells\cite{Goban:15}. Therefore, it was not possible to have precise positioning of individually trapped atoms and full control of their photon-mediated interactions. By adapting techniques developed in free-space 1D, 2D and 3D atom assemblies in optical tweezer arrays\cite{Endres:16,Barredo:16,Barredo:18}, our goal here is to achieve efficient atom assembly on the PCWs with each single atom precisely positioned with respect to the nanophotonic structures as in Figure~\ref{fig:1D2D}. In this section, we present a description of the experimental protocol for trapping single atoms in a 1D tweezer array near PCWs with our advanced apparatus. 

As shown in Figure \ref{fig:2_setup}, our apparatus consists of two vacuum glass cells named source cell (top) and science cell (down) in a top-down configuration which is parallel to the gravity direction. The experiment cycle starts with the loading of a magneto-optical trap (MOT) from background, room temperature Cs vapor in the source cell for a duration $\sim 1$ s. With $\sim 10^7$ atoms loaded into the $\sim 2$ mm  effective diameter MOT, we then perform a 10 ms polarization gradient cooling (PGC)\cite{Metcalf:99} to cool the dense atom cloud to $\sim10$ $\mu$K before transferring into a blue-detuned donut-shaped dipole trap beam which guides falling cold atoms down into the science cell. Due to the $\sim0.5$ m separation between source cell and science cell, it takes $\sim 300$ ms for cold atom freely falling from the source cell, with total delivery efficiency $\sim 20\%$ which is limited by the lifetime of atoms in the blue dipole trap.  In the science cell atoms are stopped and then cooled by PGC to a volume of $\sim (200 $ $\mu m)^3$ with temperature $\simeq 20$ $\mu$K. 

\begin{figure}[htbp]
\centering
\fbox{\includegraphics[width=\linewidth]{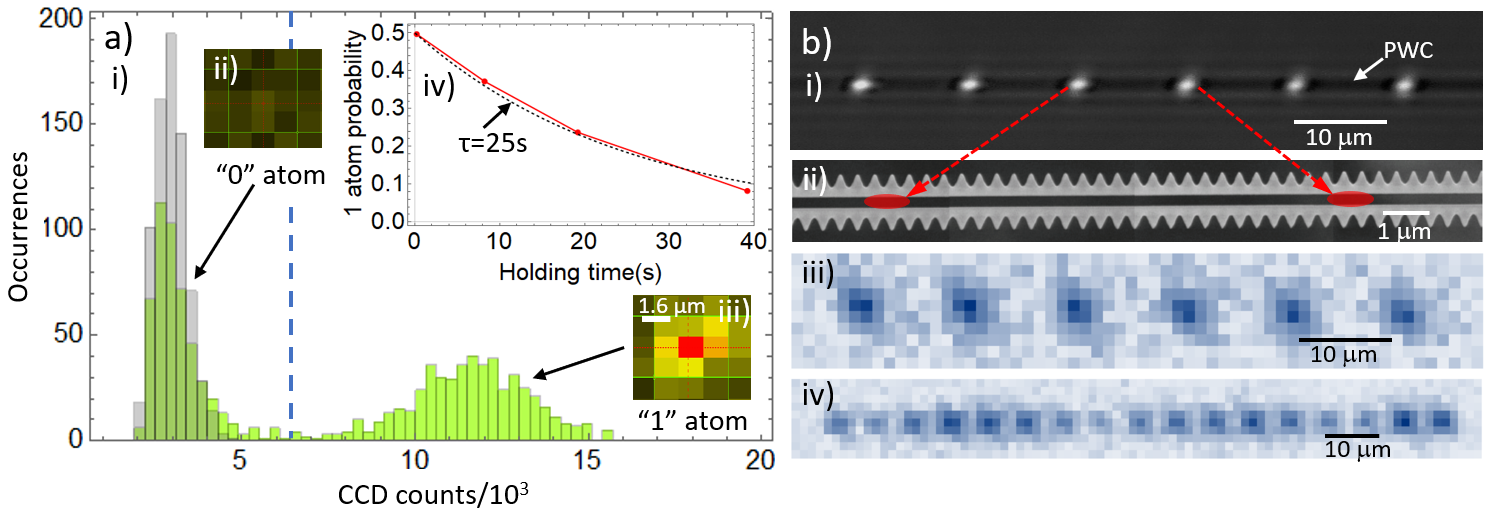}}
\caption{(a) (i) A typical histogram of fluorescence counts(green color) measured from a single site of the tweezer traps shows a discrete distribution of 0 and 1 atom loaded each time, as compared to the background(gray) with no loading. The vertical blue dashed line sets the threshold of detecting 1 atom. Inserted (ii) and (iii) indicate the typical image on EMCCD for `0' atom and `1' atom, respectively. Inserted (iv) shows the extracted probability of detecting 1 atom after different holding time of the tweezer without cooling. An exponential fit shows the average $1/e$ lifetime of a atom inside the tweezer is 25 s for vacuum cell after baking. (b) (i) Image of scattered light from tweezer spots when aligned with the APCW, collected through the same tweezer objective. (ii) An SEM image of the APCW with red ellipses indicate the separation of two neighbour tweezer spots. The size of red ellipses indicate the estimated confinement of an atom trapped with energy half the trap depth. (iii) Free-space atomic fluorescence from loading of the six tweezer sites with a 1.26 $\mu$m beam waist for 150 experimental shots 3 mm away from the chip structure. (iv) Free-space atomic fluorescence from loading of the 17 tweezer sites under same conditions in (iii).}
\label{fig:2_atom}
\end{figure}

\begin{figure}[hbp]
\centering
\fbox{\includegraphics[width=\linewidth]{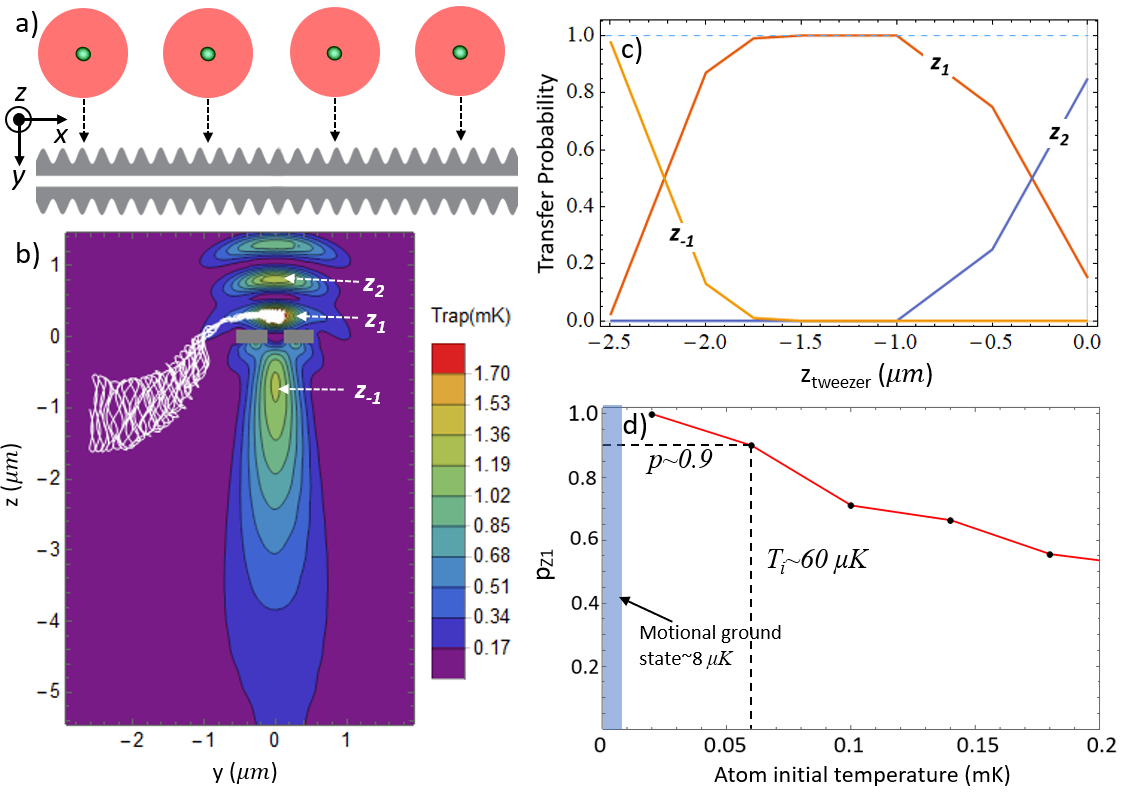}}
\caption{(a) An illustration of the scheme for transferring a free-space array of single atoms to reflective traps above PCWs. The red circles indicate the size of a tweezer with radius equal to the waist $w_0=850$ nm and green dots indicate the rms span sizes of $\sim 100$  $\mu$K trapped atoms in a 1 mK tweezer trap. (b) A COMSOL\textsuperscript{\textregistered}\cite{Comsol55} simulation of the tweezer trap when aligned on PCWs with tweezer focus position at $z=-1$ $\mu$m and polarization along x direction (out of page), with the closest three reflective traps labeled as `$z_1$', `$z_{-1}$' and `$z_2$'. The white curve shows a typical atom trajectory starting at initial tweezer trap minimum $(x,y,z)=(0,-2.5,-1)$ $\mu$m with initial temperature $\sim$ 50 $\mu$K. (c) Probability of transferring into different reflective traps at different tweezer focus positions along z direction, with atom initial temperature $T_i=20$ $\mu$K. (d) Probability of transferring into $z_1$ trap at different atom initial temperatures at 1 mK tweezer trap.}
\label{fig:2_transfer}
\end{figure}

Next, individual Cesium atoms from the PGC cloud are loaded into a linear array of optical tweezers with trap depths $U_{trap}/k_B\sim 1$ mK at a Cs magic wavelength (935.7 nm). The tweezer array is generated by sending the output of an acoustic-optical deflector (AOD) with RF-controlled spacing into a tweezer objective ($N.A. = 0.4$), forming focal spots with $1/e^2$ waist $w_0 \simeq 1.26 \pm 0.15 $ $\mu$m. After 30 ms of PGC and loading into the tweezer array, we turn off the PGC beams for 50ms to let background atoms drop away and then turn on the same PGC beams (15 MHz red detuned from D2, $F=4-5'$ transition) to illuminate the trapped atoms for 50  ms. As shown in Figure~\ref{fig:2_atom}a (i), (ii) and (iii), by binning fluorescence counts recorded on the EMCCD (Andor Camera iXon 3) around each tweezer location, we demonstrate that the distributions of fluorescence counts are well separated due to discrete loading of either 0 or 1 atom into each tweezer spot with approximately equal probability. This corresponds to the `collisional blockade' for loading atoms into tightly focused dipole traps\cite{Schlosser:01,Schlosser:02}. To determine the lifetime of a trapped atom in the optical tweezer, we measured the occurrence of `1 atom' events after different holding times, with cooling light shut off by an optical shutter. As shown in Figure~\ref{fig:2_atom}a (iv), we measured $\simeq$ 25 s for the first baked vacuum cell and 14 s for the unbaked vacuum cell (not shown). To minimize the impact of light scattering from the chip during trap loading and also the Cs atoms deposition on PCWs from high Cs densities\cite{McClung:17}, the tweezer array is loaded approximately 3 mm away from the surface of the Silicon chip. Due to the relatively long trap lifetime and flexibility afforded by external objective lenses, transport of atoms trapped in the tweezer array to near the surface of the Silicon chip along the PCW is accomplished over a programmable interval $0.02 <\Delta t < 0.1$ s by mounting the tweezer objective on a precision linear translation stage (model Physik Instrumente V-522, $20$ nm unidirectional repeatability) with motion along the $z$ direction defined in Figure \ref{fig:2_setup}. 

To investigate the efficiency for transport of single atoms in a linear array of tweezer traps, we measured the conditional survival probability $P_s$ by transporting single atoms from the loading zone to a target position near the PCW ($\Delta t \simeq 0.1$ s), holding still at target position for $\simeq 0.1$ s, and then moving back ($\Delta t \simeq 0.1$ s) to the loading zone for a second fluorescence imaging. In this measurement, a specific target position is chosen to be $\sim$ 10 $\mu$m away from the APCW (along $y$ axis) and in the APCW's $x-y$ plane, as indicated by the green dots in Figure \ref{fig:2_transfer}a. Given an initial measurement that verifies that a particular tweezer trap is loaded, we find that $P_s \simeq 0.90$ for transport from the loading zone to a target position and back to the loading zone for a second fluorescence measurement. This observation suggests that the one-way success probability for transport from loading zone to target position is $\simeq 0.95$ and the lifetime of trapped atoms at the target position near the PCW is $\geq 2$ s.

Examples of atom loading into 6 and 17 tweezer sites far from the APCW are shown in Figure~\ref{fig:2_atom}b (iii) and (iv). Figure~\ref{fig:2_atom}b (i) displays the reflection of multiple tweezer spots from the APCW with the target positions now being inside the gap of the APCW (i.e., $y=0$ and along the x-axis), albeit with no atom imaging. Beyond these initial measurements, sub-micron waists are achievable with a higher numerical aperture objective (N.A. $\sim$ 0.7) and will be discussed in Section \ref{sec:highNA}.

After transport of the 1D trapped atoms array to the target position $\sim 10$ $\mu$m from the PCW, the single atom 1D array will be further transferred into reflective traps near the dielectric surface of the PCW for strong atom-light interactions\cite{Thompson:13,Burgers:19,Kim:19}. Here, we numerically investigate a protocol for atom assembly on the PCW inspired by Refs.~\onlinecite{Thompson:13} and \onlinecite{Endres:16}, now within the setting of the APCW. As illustrated in Figure \ref{fig:2_transfer}a, moving the entire tweezer array vertically along the $y$ direction is achieved by sweeping the RF drive frequency for a second AOD with axis along $y$ orthogonal to that of the first AOD forming the original tweezer array along $x$. Figure \ref{fig:2_transfer}b shows a typical atom trajectory (white curve) from numerical simulation being successfully transferred into the so-called $z_1$ trap close to the upper surface of the APCW trap with the largest coupling rate to TE mode of the APCW\cite{Goban:15,Hood:16}. Here, the free-space initial position of the tweezer waist  is located at $(x,y,z)=(0,-2.5,-1)$ $\mu$m with the atom initial temperature $\sim$ 50 $\mu$K.

More generally, the probabilities of transferring into different reflective trap sites $\{z_i\}$ can be tuned by changing the initial tweezer focus along the z direction. This is further quantified by a Monte-Carlo simulation of atom trajectories as shown in Figure \ref{fig:2_transfer}c for tweezer waist $w_0 =850$ nm and atom initial temperature $20$ $\mu$K. The simulated probability of transferring into the $z_1$ trap is peaked at $\sim 100\%$ between $z=-1$ $\mu$m and $z=-1.7$ $\mu$m which indicates a relatively large tolerance of tweezer focus positions along $z$. Our investigations of atom trajectories via Monte-Carlo simulation show that achieving high probability transfer with ($p>0.9$) into the $z_1$ trap also requires that the atom starts initially with a temperature less than $60$ $\mu$K, as shown in Figure \ref{fig:2_transfer}c for tweezer waist $w_0=850$ nm, trap depth $U/k_B=1$ mK and focus position $z=-1.5$ $\mu$m. Absent the difficulties brought by the dielectric boundary, this can be achieved with PGC in the tweezer or Raman sideband cooling as shown in Refs. ~\onlinecite{Kaufman:12,ThompsonPRL:13}. Once atoms are transferred into $z_1$ trap, further transferring into the guided mode trap can be achieved by adiabatically turning off tweezer traps and turning on suitable guided mode traps\cite{Hung:13,Goban:14,Burgers:19,Pengthesis:19}. 

For the APCW as well as more complicated structures such as 2D photonic crystals in Figure \ref{fig:1D2D}b, we have recently proposed a quite different scheme for direct delivery with high efficiency of single atoms from free-space tweezer traps into $z_1$ traps at the surfaces of nanophotonic structures\cite{BeguinLG:20}. In this work, we exploit the rapid spatial variation of the Gouy phase for optical tweezers formed by radial Laguerre-Gauss beams to reduce the trap size in the axial direction.

\section{Towards higher optical resolution\label{sec:highNA}}

\begin{table}[htbp]
\centering
\caption{\textbf{Waist benchmark of higher N.A. objectives.} f and W.D. mean focal length and working distance, respectively.}
\begin{tabular}{ccccccc}
\hline
N.A. & f &  W.D. & Input waist & Waist in air & Waist in vacuum \\
 & (mm) & (mm) & (mm) & ($\mu$m) & ($\mu$m) \\
\hline
0.4 & 10 & 25 & 2.5 & 1.22 & 1.23\\
0.67 & 4 & 10.5 & 1.8 & 1.02 & 1.02\\
0.7 & 2 & 6 & 1.1 & 0.91 & 0.87\\
\hline
\end{tabular}
  \label{tab:waists}
\end{table}

Our current apparatus can be improved in terms of optical resolution for both imaging and smaller trap volume. It can accommodate state-of-the-art long-working distance objective lenses with $N.A.\sim0.7$. To characterize the tweezer waist under different N.A. objectives and filling ratio, we employed three different methods for a cross-check: (i) Imaging by a CCD camera and a telescope system which consists of a $N.A.=0.8$  objective and 125 mm focal length lens. (ii) A knife-edge experiment\cite{Suzaki:75} with a razor blade glued inside a fused silica cell with inner dimensions of 2 cm by 1 cm. The objective is moved on a two-dimensional motorized stage to scan along and perpendicular to the light propagation direction. The Rayleigh range and tweezer waist can be fitted from the transmitted light after the glass cell. To investigate the effect of glass cell bowing on tweezer waist, the measurement is performed when the glass cell is either at atmospheric pressure or under vacuum (below $10^{-4}$ Torr). (iii) For tweezers inside the science glass cell, the tweezer beam is scanned across a 500 nm wide and 200 nm thick uniform waveguide region inside the glass cell, and the reflection and scattering are imaged by an EMCCD. The Rayleigh range and tweezer waist can be obtained by fitting the images. The tweezer waist measured from these three different methods for the $N.A.=0.4$ objective are all around 1.25 $\mu$m and consistent within 10\%.

\begin{figure}[t]
\centering
\fbox{\includegraphics[width=\linewidth]{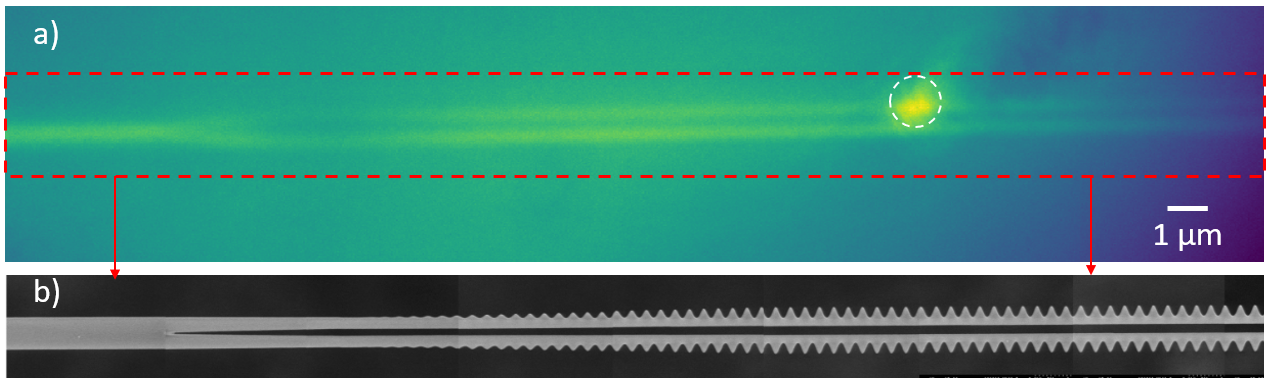}}
\caption{(a) A CCD image of APCW inside a glass cell with $N.A.=0.7$ objective under broadband illumination at 532 nm. The focused tweezer is at 852 nm with $1/e^2$ waist $w_0 \simeq 0.7 $ $\mu$m, indicated as white dashed circle with radius = 0.7 $\mu$m. The junction between the nanobeam waveguide and the APCW region is resolved. (b) A SEM image of the part of APCW under image, corresponding to the red dashed box in (a).}
\label{fig:4}
\end{figure}

Table \ref{tab:waists} shows the measured beam waist $w_0$ ($1/e^2$ radius for intensity) with the razor blade method for three different NA objectives ($N.A.=0.4$ Mitutoyo M Plan Apo NIR B 20X, $N.A.=0.67$ OptoSigma PAL-50-NIR-HR, compensated for glass thickness 1.25 mm and Mitutoyo G Plan Apo 100X, compensated for glass thickness 1 mm). The tweezer waists can reach 1.02 $\mu$m for $N.A.=0.67$ objective and sub-$\mu$m for $N.A.=0.7$ objective. It shows that the differences for tweezer waists in air and under vacuum are negligible for all three different NA objectives and thus the bowing effect of the glass cell does not significantly contribute to the tweezer aberration. 

In Figure \ref{fig:4}a, we show an image of the APCW inside a glass cell imaged with $N.A.=0.7$ objective under broadband illumination at 532 nm. The high resolution of the $N.A.=0.7$ objective allows us to resolve the gap of APCW and precise location of tweezer spot, as shown in Figure \ref{fig:4}a with the tweezer aligned on one nanobeam of the APCW (white dashed circle). Further atom trapping near PCWs in tweezer arrays focused by high resolution objectives is a work under progress.

\section{Summary and Outlook}

We have presented an advanced apparatus for the integration of atoms and nanophotonics with several significant advances, including 1) efficient free-space coupling of light to and from guided modes of PCWs with greatly improved power handling capabilities relative to our previous work\cite{Goban:14,Goban:15,Hood:16}, 2) silicate bonding of silicon chips within small glass vacuum cells thereby reducing the volume of the (bake-able) vacuum envelope of our systems from $\sim1$ liter \cite{Yu:14,Munizthesis:17} to $\sim 5$ cm$^3$ with an associated drop in pressure $>100$x, and 3) deterministic, mechanical delivery of $1D$ single-atom tweezer arrays to near an APCW. Each of these advances eliminates significant impediments present in prior experiments in the Quantum Optics Group at Caltech as explained in previous sections.

The advanced atom-nanophotonic platform that we have described can provide a foundation for realizing nanoscopic atomic lattices in $1D$ and $2D$ as envisioned in Figure \ref{fig:1D2D} and thereby for experimental explorations of these systems involving \textit{strong interactions between atoms and photons} in nanophotonic structures as described in Ref.~\onlinecite{Kimble:18,Lodahl:17}. Attaining deterministic atom arrays, in the spirit of recent worldwide advances with free-space tweezer arrays~\cite{Kaufman:12,ThompsonPRL:13,Endres:16,Barredo:16,Barredo:18}, will allow us to probe the physics of strong, photon-mediated interactions between many atoms, as well as atom-mediated interactions between photons. The versatility of dispersion-engineered nanostructures makes accessing these physical regimes possible in a single cold-atom experiment by changing the nanophotonic structures the atom interacts with.

Moreover, the compact nature of our system also lends itself to more easily deployable quantum technologies that are of growing interest in the community. Possible applications of these nanophotonic systems range from quantum communication using strong atom-photon interactions to probing unconventional quantum phase transitions and investigating quantum metrology applications by combining the functionality of nanophotonics and atoms. One example relates to ongoing investigations of the integration of nanophotonic systems such as described here with on-chip frequency combs for a compact atomic frequency standard\cite{Picque:19,Hummon:18}. The collective decay of $N$ atoms (known as Dicke superradiance) into PCWs demonstrated in Ref.~\onlinecite{Goban:15} could provide a simple, deterministic and scalable way to generate Fock states with large and fixed photon numbers, enable quantum-enhanced metrology\cite{Paulisch:19}. Apart from Quantum Information Science, an essential aspect of atom trapping near nanophotonic structures is a quantitative understanding of Casimir-Polder interactions between trapped atoms and the dielectric boundaries\cite{Hung:13,Gonzalez-Tudela:15}. We have taken a modest step toward this end in recent work\cite{Burgers:19}.

With a broad set of such objectives in mind, we have developed the advanced apparatus described in this manuscript. In terms of an outlook for closing the aforementioned gap described in the introduction between experiment and theory (e.g., Ref.~\onlinecite{Kimble:18}), there are two issues to address. The first is the development of an apparatus suited to the laboratory realization of the current cartoon depicted in Figure \ref{fig:1D2D} for dense filling of 1D and 2D PCWs with single atoms, which our advanced apparatus should be capable of achieving. However, the second challenge is that the PCW should excel beyond the current APCW in terms of coupling strength between single atoms and photons within a guided mode of the PCW. For a 1D PCW, the still current state-of-the-art is the experiment in Ref. \onlinecite{Hood:16}.

This experiment and its possible improvements are reviewed in Ref. \onlinecite{Asenjo-Garcia:17}. For previous experiments with the APCW\cite{Goban:15,Hood:16}, the ratio of the waveguide coupling rate $\Gamma_{1D}^{max}(\nu_1)$ for an atom trapped within the vacuum gap to the rate $\Gamma^{\prime}$ of atomic spontaneous decay to all other modes is $P=\Gamma_{1D}^{\max} (\nu_1)/\Gamma^{\prime} \simeq 1.4$. Here the notation is as described in Ref. \onlinecite{Hood:16}, with $\nu_1$ a resonance frequency near the dielectric band edge of the APCW. However, by moving from the APCW to a more advanced PCW that has already been fabricated and tested, namely the Slot Photonic Crystal Waveguide (SPCW) described in Ref.\onlinecite{Yuthesis:17}, we would have $P=\Gamma_{1D}^{\max} (\nu_1)/\Gamma^{\prime} \simeq 44$. For operation with the Cs frequency within the bandgap, the authors of Ref.~\onlinecite{Asenjo-Garcia:17} project the ratio $R$ of coherent `spin' exchange $J_{1D}$ to incoherent guided-mode loss $\Gamma_{1D}$ to be $R=J_{1D}/\Gamma_{1D} \simeq 20$ at a detuning of 20 GHz from the band edge, with the possibility to observe coherent spin-exchange oscillations between two proximal atoms trapped within the SPCW shown in Figure 6b in Ref. ~\onlinecite{Asenjo-Garcia:17}. To achieve such an advance would require a) the capabilities described in this manuscript for placing two tweezer-trapped atoms precisely along the SPCW, transfer from the tweezer traps to GM traps within the vacuum gap for $5\times$ increased coupling rate, and c) a next generation PCW such as the SPCW.

Apart from improved coupling in 1D PCWs, switching to 2D nanophotonic structures offers many new opportunities such as anisotropic emissions\cite{Gonzalez-Tudela:19,Yu:19}, Markovian and non-Markovian dynamics\cite{Gonzalez-Tudela:17} and topological quantum optics\cite{Perczel:20}. An example of a basic experiment can involve only a pair of atoms in two tweezer traps near the surface of the 2D photonic crystals. The anisotropic character of 2D photonic bands for photonic crystal structures described in Ref.\onlinecite{Yu:19} could be mapped by exciting one `emitter' atom and monitoring the fluorescence counts from  another neighbour `probe' atom. By varying the relative angle and distance between the `emitter' and `probe' atoms, we could measure the anisotropic emission of `emitter' atom and from that infer the spatial information of the 2D Green function as described in Ref.~\onlinecite{Asenjo-Garcia:17}.

\section*{Additional acknowledgements}
The authors thank John Hall and Jun Ye (JILA), Julien Laurat (Sorbonne University), Norma Robertson (Caltech), Keith Hulme (Starna), Jeff Gabriel (PI) and Craig Goldberg (Newport) for important discussions.

\noindent\textbf{Funding Information.} We acknowledge support from the following grants and organizations: ONR Grant No. N000141612399, ONR MURI Quantum Opto-Mechanics with Atoms and Nanostructured Diamond Grant No. N000141512761, AFOSR MURI Photonic Quantum Matter Grant No. FA95501610323, and NSF Grant No. PHY1205729, as well as the Caltech KNI.

\noindent\textbf{Conflict of Interest:} Each author declares that he has no financial/commercial Conflict of Interest.

\bibliographystyle{ChemEurJ}

\bibliography{AQT_ref}

\providecommand{\url}[1]{\texttt{#1}}
\providecommand{\urlprefix}{}
\providecommand{\foreignlanguage}[2]{#2}
\providecommand{\Capitalize}[1]{\uppercase{#1}}
\providecommand{\capitalize}[1]{\expandafter\Capitalize#1}
\providecommand{\bibliographycite}[1]{\cite{#1}}
\providecommand{\bbland}{and}
\providecommand{\bblchap}{chap.}
\providecommand{\bblchapter}{chapter}
\providecommand{\bbletal}{et~al.}
\providecommand{\bbleditors}{editors}
\providecommand{\bbleds}{eds.}
\providecommand{\bbleditor}{editor}
\providecommand{\bbled}{ed.}
\providecommand{\bbledition}{edition}
\providecommand{\bbledn}{ed.}
\providecommand{\bbleidp}{page}
\providecommand{\bbleidpp}{pages}
\providecommand{\bblerratum}{erratum}
\providecommand{\bblin}{in}
\providecommand{\bblmthesis}{Master's thesis}
\providecommand{\bblno}{no.}
\providecommand{\bblnumber}{number}
\providecommand{\bblof}{of}
\providecommand{\bblpage}{page}
\providecommand{\bblpages}{pages}
\providecommand{\bblp}{p}
\providecommand{\bblphdthesis}{Ph.D. thesis}
\providecommand{\bblpp}{pp}
\providecommand{\bbltechrep}{Tech. Rep.}
\providecommand{\bbltechreport}{Technical Report}
\providecommand{\bblvolume}{volume}
\providecommand{\bblvol}{Vol.}
\providecommand{\bbljan}{January}
\providecommand{\bblfeb}{February}
\providecommand{\bblmar}{March}
\providecommand{\bblapr}{April}
\providecommand{\bblmay}{May}
\providecommand{\bbljun}{June}
\providecommand{\bbljul}{July}
\providecommand{\bblaug}{August}
\providecommand{\bblsep}{September}
\providecommand{\bbloct}{October}
\providecommand{\bblnov}{November}
\providecommand{\bbldec}{December}
\providecommand{\bblfirst}{First}
\providecommand{\bblfirsto}{1st}
\providecommand{\bblsecond}{Second}
\providecommand{\bblsecondo}{2nd}
\providecommand{\bblthird}{Third}
\providecommand{\bblthirdo}{3rd}
\providecommand{\bblfourth}{Fourth}
\providecommand{\bblfourtho}{4th}
\providecommand{\bblfifth}{Fifth}
\providecommand{\bblfiftho}{5th}
\providecommand{\bblst}{st}
\providecommand{\bblnd}{nd}
\providecommand{\bblrd}{rd}
\providecommand{\bblth}{th}
\begin{thebibliography}{10}

\bibitem{Kimble:18}
D.~E. Chang, J.~S. Douglas, A.~Gonz\'alez-Tudela, C.-L. Hung, H.~J. Kimble,
  \emph{Rev. Mod. Phys.} \textbf{2018}, \emph{90}, 031002.

\bibitem{Lodahl:17}
P.~Lodahl, S.~Mahmoodian, S.~Stobbe, A.~Rauschenbeutel, P.~Schneeweiss,
  J.~Volz, H.~Pichler, P.~Zoller, \emph{Nature} \textbf{2017}, 473--490.

\bibitem{Gonzalez-Tudela:15}
A.~González-Tudela, C.-L. Hung, D.~E. Chang, J.~I. Cirac, H.~J. Kimble,
  \emph{Nature Photonics} \textbf{2015}, \emph{9}, 320--325.

\bibitem{Douglas:15}
J.~S. Douglas, H.~Habibian, C.-L. Hung, A.~V. Gorshkov, H.~J. Kimble, D.~E.
  Chang, \emph{Nature Photonics} \textbf{2015}, \emph{9}, 326--331.

\bibitem{Yu:19}
S.-P. Yu, J.~A. Muniz, C.-L. Hung, H.~J. Kimble, \emph{Proceedings of the
  National Academy of Sciences} \textbf{2019}, \emph{116}, 12743--12751.

\bibitem{Kimble:08}
H.~J. Kimble, \emph{Nature} \textbf{2008}, \emph{453}, 1023--1030.

\bibitem{Hung:16}
C.-L. Hung, A.~Gonz{\'a}lez-Tudela, J.~I. Cirac, H.~J. Kimble,
  \emph{Proceedings of the National Academy of Sciences} \textbf{2016},
  \emph{113}, E4946--E4955.

\bibitem{Covey:19}
J.~P. Covey, A.~Sipahigil, S.~Szoke, N.~Sinclair, M.~Endres, O.~Painter,
  \emph{Phys. Rev. Applied} \textbf{2019}, \emph{11}, 034044.

\bibitem{Sipahigil:16}
A.~Sipahigil, R.~E. Evans, D.~D. Sukachev, M.~J. Burek, J.~Borregaard, M.~K.
  Bhaskar, C.~T. Nguyen, J.~L. Pacheco, H.~A. Atikian, C.~Meuwly, R.~M.
  Camacho, F.~Jelezko, E.~Bielejec, H.~Park, M.~Lon{\v c}ar, M.~D. Lukin,
  \emph{Science} \textbf{2016}, \emph{354}, 847--850.

\bibitem{Dibos:18}
A.~M. Dibos, M.~Raha, C.~M. Phenicie, J.~D. Thompson, \emph{Phys. Rev. Lett.}
  \textbf{2018}, \emph{120}, 243601.

\bibitem{Zhong:18}
T.~Zhong, J.~M. Kindem, J.~G. Bartholomew, J.~Rochman, I.~Craiciu, V.~Verma,
  S.~W. Nam, F.~Marsili, M.~D. Shaw, A.~D. Beyer, A.~Faraon, \emph{Phys. Rev.
  Lett.} \textbf{2018}, \emph{121}, 183603.

\bibitem{Evans:20}
R.~E. Evans, M.~K. Bhaskar, D.~D. Sukachev, C.~T. Nguyen, A.~Sipahigil, M.~J.
  Burek, B.~Machielse, G.~H. Zhang, A.~S. Zibrov, E.~Bielejec, H.~Park,
  M.~Lon{\v c}ar, M.~D. Lukin, \emph{Science} \textbf{2018}, \emph{362},
  662--665.

\bibitem{Thompson:13}
J.~D. Thompson, T.~Tiecke, N.~P. de~Leon, J.~Feist, A.~Akimov, M.~Gullans,
  A.~S. Zibrov, V.~Vuleti{\'c}, M.~D. Lukin, \emph{Science} \textbf{2013},
  \emph{340}, 1202--1205.

\bibitem{Tiecke:14}
T.~Tiecke, J.~D. Thompson, N.~P. de~Leon, L.~Liu, V.~Vuleti{\'c}, M.~D. Lukin,
  \emph{Nature} \textbf{2014}, \emph{508}, 241.

\bibitem{Goban:14}
A.~Goban, C.-L. Hung, S.-P. Yu, J.~D. Hood, J.~A. Muniz, J.~H. Lee, M.~J.
  Martin, A.~McClung, K.~Choi, D.~E. Chang, H.~J. Kimble, \emph{Nat. Commun.}
  \textbf{2014}, \emph{5}, 3808.

\bibitem{Goban:15}
A.~Goban, C.-L. Hung, J.~D. Hood, S.-P. Yu, J.~A. Muniz, O.~Painter, H.~J.
  Kimble, \emph{Phys. Rev. Lett.} \textbf{2015}, \emph{115}, 063601.

\bibitem{Hood:16}
J.~D. Hood, A.~Goban, A.~Asenjo-Garcia, M.~Lu, S.-P. Yu, D.~E. Chang, H.~J.
  Kimble, \emph{Proc. Natl. Acad. Sci. U.S.A.} \textbf{2016}, \emph{113},
  10507--10512.

\bibitem{Samutpraphoot:20}
P.~Samutpraphoot, T.~Dordević, P.~L. Ocola, H.~Bernien, C.~Senko, V.~Vuletić,
  M.~D. Lukin, \emph{Phys. Rev. Lett.} \textbf{2020}, \emph{124}, 063602.

\bibitem{Yu:14}
S.-P. Yu, J.~D. Hood, J.~A. Muniz, M.~J. Martin, R.~Norte, C.-L. Hung, S.~M.
  Meenehan, J.~D. Cohen, O.~Painter, H.~J. Kimble, \emph{Applied Physics
  Letters} \textbf{2014}, \emph{104}, 111103.

\bibitem{Hung:13}
C.-L. Hung, S.~M. Meenehan, D.~E. Chang, O.~Painter, H.~J. Kimble, \emph{New
  Journal of Physics} \textbf{2013}, \emph{15}, 083026.

\bibitem{Burgers:19}
A.~P. Burgers, L.~S. Peng, J.~A. Muniz, A.~C. McClung, M.~J. Martin, H.~J.
  Kimble, \emph{Proceedings of the National Academy of Sciences} \textbf{2019},
  \emph{116}, 456--465.

\bibitem{Yuthesis:17}
S.~P. Yu, \bblphdthesis{}, California Institute of Technology, \textbf{2017}.

\bibitem{Kaufman:12}
A.~M. Kaufman, B.~J. Lester, C.~A. Regal, \emph{Phys. Rev. X} \textbf{2012},
  \emph{2}, 041014.

\bibitem{ThompsonPRL:13}
J.~D. Thompson, T.~G. Tiecke, A.~S. Zibrov, V.~Vuleti\ifmmode~\acute{c}\else
  \'{c}\fi{}, M.~D. Lukin, \emph{Phys. Rev. Lett.} \textbf{2013}, \emph{110},
  133001.

\bibitem{Endres:16}
M.~Endres, H.~Bernien, A.~Keesling, H.~Levine, E.~R. Anschuetz, A.~Krajenbrink,
  C.~Senko, V.~Vuletic, M.~Greiner, M.~D. Lukin, \emph{Science} \textbf{2016},
  3752.

\bibitem{Barredo:16}
D.~Barredo, S.~de~L{\'e}s{\'e}leuc, V.~Lienhard, T.~Lahaye, A.~Browaeys,
  \emph{Science} \textbf{2016}, \emph{354}, 1021--1023.

\bibitem{Munizthesis:17}
J.~A.~M. Silva, \bblphdthesis{}, California Institute of Technology,
  \textbf{2017}.

\bibitem{Beguin:20}
J.-B. B\'{e}guin, A.~P. Burgers, X.~Luan, Z.~Qin, S.~P. Yu, H.~J. Kimble,
  \emph{Optica} \textbf{2020}, \emph{7}, 1--2.

\bibitem{McClung:17}
A.~C. McClung, \bblphdthesis{}, California Institute of Technology,
  \textbf{2017}.

\bibitem{Anderson95}
M.~H. Anderson, J.~R. Ensher, M.~R. Matthews, C.~E. Wieman, E.~A. Cornell,
  \emph{Science} \textbf{1995}, \emph{269}, 198--201.

\bibitem{Veggel:14}
A.-M.~A. van Veggel, C.~J. Killow, \emph{Advanced Optical Technologies}
  \textbf{2014}, \emph{3}, 293--307.

\bibitem{Gwo:01}
D.-H. Gwo, \emph{Ultra precision and reliable bonding method}, \textbf{U.S.
  Patent 6 284 085, Sep. 2001}.

\bibitem{Mangano:17}
V.~Mangano, A.~A. van Veggel, R.~Douglas, J.~Faller, A.~Grant, J.~Hough,
  S.~Rowan, \emph{Opt. Express} \textbf{2017}, \emph{25}, 3196--3213.

\bibitem{Starna}
\emph{Starna Cells Inc.}, \url{http://www.starnacells.com/}, [Online; accessed
  06-Jan-2020].

\bibitem{Ye:08}
J.~Ye, H.~J. Kimble, H.~Katori, \emph{Science} \textbf{2008}, \emph{320},
  1734--1738.

\bibitem{Vetsch:10}
E.~Vetsch, D.~Reitz, G.~Sagu{\'e}, R.~Schmidt, S.~Dawkins, A.~Rauschenbeutel,
  \emph{Phys. Rev. Lett.} \textbf{2010}, \emph{104}, 203603.

\bibitem{Goban:12}
A.~Goban, K.~S. Choi, D.~J. Alton, D.~Ding, C.~Lacro\^ute, M.~Pototschnig,
  T.~Thiele, N.~P. Stern, H.~J. Kimble, \emph{Phys. Rev. Lett.} \textbf{2012},
  \emph{109}, 033603.

\bibitem{Cohen:13}
J.~D. Cohen, S.~M. Meenehan, O.~Painter, \emph{Opt. Express} \textbf{2013},
  \emph{21}, 11227--11236.

\bibitem{Lumerical}
Lumerical\textsuperscript{\textregistered}, \emph{{Lumerical FDTD Reference
  Manual}}, \url{https://www.lumerical.com/products/}, \textbf{2020}.

\bibitem{Barredo:18}
D.~Barredo, V.~Lienhard, S.~de~Léséleuc, T.~Lahaye, A.~Browaeys,
  \emph{Nature} \textbf{2018}, \emph{561}, 79--82.

\bibitem{Metcalf:99}
H.~J. Metcalf, P.~van~der Straten, \emph{Laser Cooling and Trapping},
  Springer-Verlag, New York, \textbf{1999}.

\bibitem{Comsol55}
COMSOL\textsuperscript{\textregistered}, \emph{{COMSOL Multiphysics Reference
  Manual, version 5.5}}, \url{www.comsol.com}, \textbf{2020}.

\bibitem{Schlosser:01}
N.~Schlosser, G.~Reymond, I.~Protsenko, P.~Grangier, \emph{Nature}
  \textbf{2001}, \emph{411}, 1024--1027.

\bibitem{Schlosser:02}
N.~Schlosser, G.~Reymond, P.~Grangier, \emph{Phys. Rev. Lett.} \textbf{2002},
  \emph{89}, 023005.

\bibitem{Kim:19}
M.~E. Kim, T.-H. Chang, B.~M. Fields, C.-A. Chen, C.-L. Hung, \emph{Nature
  Communications} \textbf{2019}, \emph{10}, 1647.

\bibitem{Pengthesis:19}
L.~S. Peng, \bblphdthesis{}, California Institute of Technology, \textbf{2019}.

\bibitem{BeguinLG:20}
J.-B. B\'{e}guin, J.~Laurat, X.~Luan, A.~P. Burgers, Z.~Qin, H.~J. Kimble,
  \emph{arXiv:2001.11498v2} \textbf{2020}.

\bibitem{Suzaki:75}
Y.~Suzaki, A.~Tachibana, \emph{Appl. Opt.} \textbf{1975}, \emph{14},
  2809--2810.

\bibitem{Picque:19}
N.~Picqu\'{e}, T.~W. Hänsch, \emph{Nature Photonics} \textbf{2019}, \emph{13},
  146--157.

\bibitem{Hummon:18}
M.~T. Hummon, S.~Kang, D.~G. Bopp, Q.~Li, D.~A. Westly, S.~Kim, C.~Fredrick,
  S.~A. Diddams, K.~Srinivasan, V.~Aksyuk, J.~E. Kitching, \emph{Optica}
  \textbf{2018}, \emph{5}, 443--449.

\bibitem{Paulisch:19}
V.~Paulisch, M.~Perarnau-Llobet, A.~Gonz\'alez-Tudela, J.~I. Cirac, \emph{Phys.
  Rev. A} \textbf{2019}, \emph{99}, 043807.

\bibitem{Asenjo-Garcia:17}
A.~Asenjo-Garcia, J.~D. Hood, D.~E. Chang, H.~J. Kimble, \emph{Phys. Rev. A}
  \textbf{2017}, \emph{95}, 033818.

\bibitem{Gonzalez-Tudela:19}
A.~González-Tudela, F.~Galve, \emph{ACS Photonics} \textbf{2019}, \emph{6},
  221--229.

\bibitem{Gonzalez-Tudela:17}
A.~Gonz\'alez-Tudela, J.~I. Cirac, \emph{Phys. Rev. A} \textbf{2017},
  \emph{96}, 043811.

\bibitem{Perczel:20}
J.~Perczel, J.~Borregaard, D.~E. Chang, S.~F. Yelin, M.~D. Lukin, \emph{Phys.
  Rev. Lett.} \textbf{2020}, \emph{124}, 083603.

\end{thebibliography}

\section{Supplementary \label{sec:supp}}

\subsection{PI stage magnetic field}
\begin{figure}[htbp]
\centering
\fbox{\includegraphics[width=0.7\linewidth]{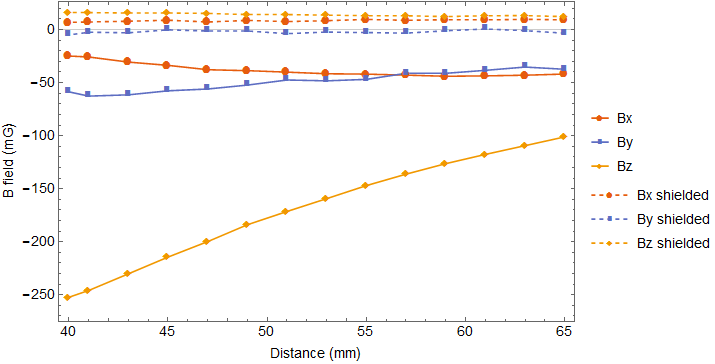}}
\caption{Measured DC magnetic field produced by PI stage. Z is along the stage translation direction.}
\label{figS:Bfield}
\end{figure}

The motor technology for the PI stage (voice coil) produces both DC and AC magnetic fields. The DC magnetic field in three directions is measured by a magnetometer as shown in Figure \ref{figS:Bfield}. The measured total magnetic field produced by the PI stage is around 250 mG at a distance of 6 cm from atoms in the science cell. The magnetic field is zeroed near the PGC cloud position by three pairs of Helmholtz coils and further suppression of magnetic field by a factor of $\sim$ 10 is also achieved by shielding the PI stage with a mu-metal box, as shown in dashed lines in Figure \ref{figS:Bfield}.

\end{document}